
\input phyzzx
\FRONTPAGE
\line{\hfill BROWN-HET-934}
\line{\hfill DAMTP 94-5}
\line{\hfill March 1994}
\vskip1.0truein
\titlestyle{\bf FORMATION OF TOPOLOGICAL DEFECTS IN A SECOND ORDER PHASE
TRANSITION}
\vskip1.0in
\author{Robert H. Brandenberger$^{1)}$ and Anne-Christine Davis$^{2)}$}
\vskip.25in
\centerline{$^{1)}$ {\it Physics Department, Brown University, Providence, RI
02912, USA}}
\centerline{$^{2)}$ {\it Department of Applied Mathematics and Theoretical
Physics}}
\centerline{\it and Kings College, University of Cambridge, Cambridge CB3 9EW,
U.K.}
\vskip2.0in
\centerline{Abstract}
\par
The classical evolution equations of the Abelian Higgs model are studied at
temperatures below the Ginsburg temperature of a phase transition which is
assumed to be second order.  It is shown that the initial thermal fluctuations
provide a domain structure which is stable against late time fluctuations.
This result lends support to the Kibble mechanism for the formation of
topological defects.

\endpage
\chapter{Introduction}

Topological defects$^{1)}$ are playing an increasing role in various branches
of physics.  In particular, cosmic strings and global textures give rise to
attractive scenarios  for the origin of structure in the early Universe (for
recent reviews see e.g., Refs. 2 and 3).  It is therefore important to obtain a
detailed understanding of the rate of formation of topological defects in phase
transitions from a hot  symmetric phase to a cold phase with broken symmetry.

The original mechanism of a defect formation is due to Kibble$^{1)}$.  He
argued that at the phase transition, in any theory which admits topological
defects, a network of such defects with correlation length (i.e., typical
separation) $\xi$ will be frozen in at the Ginsburg temperature $T_G$.  Here,
$\xi$ is the correlation length at $t_G$,  the cosmic time corresponding to
temperature $T_G$.

Starting point of Kibble's argument was the assumption that on scales larger
than $\xi$, the orientation of the order parameter in the vacuum manifold is
random, that the order parameter smoothly interpolates between these random
values, and that there thus is a finite probability to have nontrivial winding.
 This probability depends on the topology of the vacuum manifold and has been
calculated in several interesting cases in Ref. 4.

The Kibble argument has been widely used in cosmology.  For example, it has
been used to generate the initial string configurations$^{5)}$ for cosmic
string evolution studies, to calculate the abundance of magnetic
monopoles$^{6)}$, and to justify the occurrence of textures$^{7)}$ in models
with nonvanishing $\pi_3 ({\cal M})$, ${\cal M}$ being the vacuum manifold.
Another application of the Kibble mechanism is the prediction of vortices in a
pressure quench of superfluid helium$^{8)}$.

The two main assumptions of the Kibble mechanism are that the order parameter
takes on random values in ${\cal M}$ on scales larger than $\xi$, and that it
smoothly interpolates between its values at different points in space (the
``geodesic rule'').  Recently$^{9)}$, the validity of the geodesic rule has
been challenged, in particular for gauge theories.  Since the energy density in
gauge theories is proportional to $(D_\mu \phi)^2$ and not $(\partial_\mu
\phi)^2$ (where $D_\mu$ is the covariant derivative operator), the order
parameter $\phi$ need not interpolate smoothly in order to minimize the
gradient energy.  Hence, it has become important to investigate the validity of
the Kibble mechanism more carefully.

For theories with a global symmetry, the geodesic rule is well justified.  In
this case, the Kibble mechanism has been tested both in numerical
simulations$^{10)}$ and in the laboratory$^{11,12)}$.

For theories with a local symmetry the situation is less clear.  In models with
a first order phase transition, a recent analysis$^{13)}$ of the dynamics of
the classical fields has provided strong support for the Kibble mechanism.  The
main idea of the analysis, however, hinged on the phase transition proceeding
via the nucleation and subsequent collision of bubbles of the broken symmetry
phase.  The methods are therefore not directly applicable to models with a
second order phase transition.

In this letter we investigate the solution of the classical equations of motion
for an Abelian Higgs model with a second order phase transition.  We consider a
field configuration set up by thermal fluctuations at the Ginsburg temperature
and study its stability against thermal fluctuations present at later times.
We conclude that the initial domain structure is preserved, although naturally
the amplitude of the order parameter increases.  Our results lend support to
the hypothesis that the Kibble mechanism applies also to gauge theory defects
produced in a second order transition.

\chapter{System and Basic Equations}

As a toy model we consider the Abelian Higgs model with a complex scalar field
$\phi$ and a U(1) gauge connection $A_\mu$. Its Lagrangean is
$$
{\cal L} = (D_\mu \phi)^\dagger D^\mu \phi - V (\phi) - {1\over 4} F_{\mu\nu}
F^{\mu\nu} \eqno\eq
$$
where
$$
D_\mu = \partial_\mu - ie A_\mu \eqno\eq
$$
is the covariant derivative with gauge coupling constant $e$, $F_{\mu\nu}$ is
the field strength tensor, and $V(\phi)$ is the potential for $\phi$.  The
potential is chosen such that phase transition is second order.

In Lorentz gauge
$$
\partial_\mu A^\mu = 0 \eqno\eq
$$
the equations of motion for $\phi$ and $A_\mu$ become
$$
( \partial_\mu \partial^\mu -  2 i e A_\mu \partial^\mu - e^2 A_\mu A^\mu) \phi
+ 2 \, {\partial V\over{\partial |\phi |^2}} \, \phi = 0 \eqno\eq
$$
and
$$
\partial_\mu \partial^\mu A_{\nu} - 2 e^2 A_{\nu}  | \phi |^2 = - i e \phi^\ast
\vec \partial_{\nu} \phi \, . \eqno\eq
$$
It is convenient to separate Eq. (2.4) into equations for the amplitude $\rho$
and phase $\alpha$ of $\phi$.  Inserting
$$
\phi = \rho e^{i \alpha} \eqno\eq
$$
into (2.4) we obtain
$$
\partial^2 \rho - ( \partial \alpha - e A)^2 \rho - e^2 A^2 \rho + 2 {\partial
V\over{\partial \rho^2}} \rho = 0 \eqno\eq
$$
and
$$
\partial^2 \alpha + 2 (\partial^\mu \alpha - e A^\mu) \partial_\mu \rho {1\over
\rho} = 0 \, . \eqno\eq
$$

We wish to study the evolution of the solutions of the coupled system (2.5),
(2.7) and (2.8) of differential equations during a second order phase
transition.  We assume the following form for the finite temperature effective
potential
$$
V_T (\phi) = {1\over 4} \lambda |\phi |^4 - {1\over 2} (\lambda \eta^2 - \tilde
\lambda T^2) |\phi |^2 + {1\over 4} \lambda \eta^4 \, . \eqno\eq
$$
Here, $\lambda$ is the self-coupling constant of $\phi$, $\tilde \lambda$ is a
coupling constant determined by the graphs which dominate the one loop
effective potential, and $\eta$ is the value of $| \phi |$ in the vacuum
manifold (the scale of symmetry breaking).  In the following we take $\tilde
\lambda \simeq \lambda$.  $T$ is the temperature.

The phase transition proceeds by spinodal decomposition$^{14)}$.  Due to
thermal fluctuations in the initial state, the phase $\alpha$ will take on
random values on scales larger than the correlation length $\xi$.  The
temperature dependence of $\xi$ can be obtained by equating spatial gradient
energy required to set up the domains and potential energy gain by having
$\phi$ deviate from 0.  For temperatures below the critical temperature
$$
T_c = \lambda^{1/2} \tilde \lambda^{-1/2} \eta \simeq \eta \, , \eqno\eq
$$
the result for $\xi (T)$ is$^{1)}$
$$
\xi (T) \simeq \lambda^{-1/2} \eta^{-1} \left( 1 - \left({T\over T_c} \right)^2
\right)^{-1/2} \, . \eqno\eq
$$
As long as the temperature is higher than the Ginsburg temperature $T_G$, these
domains are unstable to thermal fluctuations.  The value of $T_G < T_c$ is
determined by equating thermal energy with the energy of the domains determined
 above.  The result is
$$
\xi (T_G) \simeq \lambda^{-1} T^{-1}_G \simeq \lambda^{-1} \eta^{-1} \, .
\eqno\eq
$$
These qualitative arguments suggest that below $T_G$, the domain structure is
frozen in.  Since the winding number $n_\gamma$ around a closed curve $\gamma$
is
$$
n_\gamma = - {i\over{4 \pi}} \oint \, {{\phi^\ast {^\leftrightarrow}
\partial_\mu \phi}\over{| \phi |^2 }} \, ds^\mu \, , \eqno\eq
$$
the random distribution of phases will induce a finite probability to have a
topological defect (in our case a cosmic string) in any correlation volume $\xi
(T_G)^3$.

To reach the final conclusion, we implicitly made use of the ``geodesic rule''
which says that along a curve $\gamma$ connecting two domains, the phase
$\alpha$ will smoothly interpolate between the values of $\alpha$ in the two
pluses.  This rule can be justified in a theory with global symmetry (since it
comes from minimizing the spatial gradient energy), but not in a gauge theory
(since the gauge fields can compensate large scalar field gradients).

In previous work$^{13)}$, we studied defect formation in a gauge theory with a
first order phase transition.  In this case, the phase transition proceeds by
nucleation of true vacuum bubbles ($| \phi | \sim \eta$, $\alpha$ constant) in
a surrounding sea of false vacuum $(\phi = 0)$.  Using the equations of motion
to study the evolution of $\phi$ when bubbles collide, we were able to show
that the geodesic rule is dynamically realized.

In the case of a second order transition we must use rather different methods.
We will again argue that the geodesic rule is dynamically realized.  The basic
logic is as follows: For $T > T_G$, thermal fluctuations will dominate the
field configurations for $\phi$ and $A_\mu$.  We can picture all fields as
superpositions of plane waves with uncorrelated phases.  Fluctuations with
wavelength $\xi (T)$ will dominate.  Those with $\xi < \xi (T)$ are
energetically suppressed while those with $\xi > \xi (T)$ are phase space
suppressed.

We will therefore study an initial field configuration with typical wavelength
$\xi (T)$ for which the geodesic rule is obviously satisfied.  We shall  argue
-- using the dynamical equations (2.5), (2.7) and (2.8) -- that the induced
domain structure is stable for $T \leq T_G$: the change in the phase induced by
the dynamics for $T < T_G$ is subdominant, and thermal fluctuations at $T =
T_G$ do not lead to a large perturbation from the evolution of the amplitude
$\rho$ in the absence of fluctuations.

Note that our analysis is done in a particular gauge.  However, the final
conclusion concerning the nonvanishing probability for nonzero winding number
is gauge independent.

\chapter{Perturbation Analysis}

In the following we study the initial stages of spinodal decomposition when
$\rho \ll \eta$.  We consider as initial field configuration at $T = T_G$ a
plane wave
$\rho_0 (\underline{x}), \, \alpha_0 (\underline{x}), \, A_0 (\underline{x})$
with wave number $k$ (the phases of the three fields are arbitrary).  In the
absence of fluctuations this background configuration will evolve smoothly in
time preserving the initial domain structure and winding number.

Our main goal is to study the stability of this configuration (and its winding
number) against thermal fluctuations.  The coupling to thermal fluctuations can
be modelled$^{15)}$ by introducing a second scalar field $\psi$ which is
assumed to remain in thermal equilibrium and which is coupled to $\phi$ via the
interaction Lagrangean
$$
{\cal L}_I = {1\over 2} g \psi^2 | \phi |^2 \, , \eqno\eq
$$
leading to a source term $-g \psi^2 \rho$ on the right hand side of (2.7).

The effects of thermal fluctuations are analyzed to first order in a
perturbative expansion about $(\rho_0, \alpha_0, A_0)$.  The total field
configurations are
$$
\eqalign{ \rho (\underline{x}, t) & = \rho_0 (\underline{x}, t) + \rho_I
(\underline{x} , t) \cr
\alpha (\underline{x}, t) & = \alpha_0 (\underline{x}, t) + \alpha_I
(\underline{x}, t) \cr
A (\underline{x}, t) & = A_0 (\underline{x} , t) + A_I (\underline{x}, t) \, ,
}
\eqno\eq
$$
and the equations of motion are expanded to first order in $\rho_I, \,
\alpha_I, \,  A_I$ and $g$.  All perturbations are induced by the coupling to
the thermal bath and hence will be of order $g$.  Our goal is to show that the
perturbations remain smaller than the background solution provided that
$k^{-1}$ corresponds to the Ginsburg length $\xi (T_G)$.

The perturbation equations are
$$
\eqalign{
(\partial^2_t - \nabla^2) \rho_I & - \left[ \dot \alpha^2_0 - (\nabla
\alpha_0)^2 - 2 \, {\partial V\over{\partial \rho^2}} + 2 {\partial \, \partial
V\over{\partial \rho \, \partial \rho^2}} \, \rho_0 \right] \rho_I \cr
& = \left[ - g \psi^2 + \dot \alpha_0 \dot \alpha_I - \nabla \alpha_0 \, \nabla
\alpha_I \right] \, \rho_0 \cr} \eqno\eq
$$
and
$$
( \partial^2_t - \nabla^2) \alpha_I + 2 \partial^\mu \alpha_I \partial_\mu
\rho_0 {1\over{\rho_0}} + 2 \partial^\mu \alpha_0 \partial_\mu \rho_I {1\over
\rho_0} - 2 \partial^\mu \alpha_0 \partial_\mu \rho_0 {\rho_I\over{\rho^2_0}} =
0 \, , \eqno\eq
$$
where the gauge fields have been set to zero to simplify the equations (they
will be included later).  The perturbations vanish at the initial
time $t = t_G$.

For $k < \lambda^{1/2} \eta = k_{crit}$ the background equation for $\rho_0$
has an exponential instability
$$
\partial^2_t \rho_0 \simeq m^2_I \rho_0 \simeq (\lambda \eta^2 - k^2 \alpha_0)
\rho_0 \simeq \lambda \eta^2 \rho_0 \eqno\eq
$$
with
$$
\rho_0 (\underline{x}, t) \sim e^{m_I (t-t_G)} , \, m_I \simeq \lambda^{1/2}
\eta \, . \eqno\eq
$$
The solution for the phase $\alpha_0$ can be taken to be time independent with
spatial gradient proportional to $k$.

The basic logic of our analysis is as follows: we first analyze the equation of
motion (3.3) for $\rho_I$ under the assumption that the length scale of
fluctuations in $\alpha_I$ is the same as for those in $\rho_I$.  Using the
Green's function method it can be shown that
$$
{\rho_I\over \rho_0} < 1 \>\>\> {\rm for} \>\>\> k_p < k_{crit} \, , \eqno\eq
$$
where $k_p$ is the wave number for the perturbation.  Next, we consider the
equation of motion (3.4) for $\alpha_I$ and show that modes of $\alpha_I$ do
not grow in time for wavenumbers larger than $k$.  This shows that the
assumption about the length scale of fluctuations in $\alpha_I$ in the first
step of the analysis was self-consistent.  In a final step we demonstrate that
the prescence of gauge fields does not invalidate the previous considerations.
  Now the details:

Picking out the dominant terms based on the above considerations, Eq. (3.3)
becomes
$$
( \partial^2_t - \nabla^2) \rho_I - m^2_I \rho_I = [ - g \psi^2 - \nabla
\alpha_0 \nabla \alpha_I ] \rho_0 \simeq - g \psi^2 \rho_0 \eqno\eq
$$
with the same $m_I$ as in the background equation (3.5).  This equation can be
solved using the Green function method, i.e.,
$$
\rho_I (\underline{x}, t) = - g \int^t_0 d  \tau d^3 y G_{ret} \, (t - \tau,
\underline{x} - \underline{y} ) \psi^2 (\tau, \underline{y}) \rho_0 (\tau,
\underline{y} ) \, , \eqno\eq
$$
where $G_{ret}$ is the retarded Green function.  A new time variable has been
chosen such that $t=0$ is the onset of spinodal decomposition when $\rho_I =
0$.

In order to estimate the right hand side of (3.9) we can replace $\rho_0 (\tau,
\underline{y})$ by its upper bound
$$
\rho_0 (\tau, \underline{y}) < {\cal A}_0 e^{m_I \tau} \, , \eqno\eq
$$
where ${\cal A}_0$ is the initial amplitude.  If the phase transition is rapid,
we can drop the $\tau$ dependence of the thermal source $\psi^2 (\tau,
\underline{y})$.  Inserting the form of the retarded Green function
$$
G_{ret} (x) = - {1\over{(2 \pi)^4}} \int d^4 p \, {e^{-ipx}\over{(p_0 + i
\varepsilon)^2 - \underline{p}^2 + m_I^2}} \eqno\eq
$$
we obtain
$$
\rho_I (t, \underline{x}) \sim {g {\cal A}_0\over{(2 \pi)^4}} \int d^3 p e^{i p
\cdot x} \tilde \psi^2 (\underline{p}) \int^\infty_{- \infty} d p_0 e^{-i p_0
t} \int^t_0 d \tau \, {e^{(m_I + i p_0) \tau} \over{(p_0 + i \varepsilon)^2 -
\underline{p}^2 + m_I^2 }} \, . \eqno\eq
$$
The $\tau$ integral can be done explicitly.  At large $p_0$, the resulting
integral scales as $p_0^{-3}$.  Hence, the integral over $p_0$ can be written
as an integral over a closed contour and solved using the method of residues.
It is easy to consider the cases $| p | < m_I$ and $| p | > m_I$ separately.
In both cases, the integral is bounded by $\pi / m_I^2$.  Hence, our estimate
for (3.12) is
$$
\rho_I (t , \underline{x}) \sim {g\over{2 \pi}}  {\cal A}_0 e^{m_I t}
{\pi\over{m^2_I}} \psi^2 (\underline{x}) \, . \eqno\eq
$$

The amplitude of $\psi^2 (\underline{x})$ is determined by thermal equilibrium.
 If $\psi$ is a self interacting scalar field with self coupling constant of
the order $1$, then
$$
\psi^2 (\underline{x}) \sim T^2 \, . \eqno\eq
$$
Hence, the ratio of $\rho_I$ to $\rho_0$ is
$$
{\rho_I\over \rho_0} \sim g {T^2\over{m^2_I}} \, . \eqno\eq
$$
At the onset of spinodal decomposition
$$
T = T_G < T_c = \lambda^{1/2} \tilde \lambda^{-1/2} \eta =  \lambda^{1/2}
g^{-1/2} \eta \eqno\eq
$$
(see (2.10)), and hence
$$
{\rho_I\over \rho_0} < 1 \, . \eqno\eq
$$

We conclude that thermal fluctuations do not dominate the evolution of $\rho$
for $T < T_c$ (they do dominate at higher temperatures!)  However, the
derivation makes use of two key assumptions which must be shown to be at least
self consistent with the above analysis.  Firstly, it is assumed that the wave
number of fluctuations in $\alpha_I$ are smaller than $k$.  Otherwise, the
second source term in (3.8) (proportional to $\nabla \alpha_0 \nabla \alpha_I$)
might dominate.  Secondly, it must be shown that the presence of gauge fields
does not disrupt the smooth evolution of $\rho$ and $\alpha$.

First, we analyze the evolution of $\alpha (t, \underline{x})$.  The initial
field configuration $\rho_0 (0,  \underline{x}), \, \alpha_0 (0,
\underline{x})$ is assumed to have nonvanishing winding number along a given
circle $C$ in space.  For winding number 1, the phase $\alpha_0 (0,
\underline{x})$ increases smoothly from $0$ to $2 \pi$ along $C$.  The only way
the winding can disappear is for a discontinuity in $\alpha (\underline{x})$ to
develop.  This is only possible if at some point along $C$ the modulus $\rho$
vanishes.  As shown above this is not possible as long as $\alpha_I$ does not
develop small wavelength excitations.

We thus analyze the equation (3.4) for $\alpha_I$ under the assumption that
$\rho_I / \rho_0 < 1$ in which case the last term on the left hand side of the
equation is negligible.  We approximate the second term as $2 i k_\mu
\partial^\mu \alpha_I$.  In Fourier space, the resulting equation is
$$
( \partial^2_t + {k}^2_I - 2 k \cdot k_I ) \tilde \alpha_I = 2 k \cdot k_I \,
{\tilde \rho_I\over \rho_I} \eqno\eq
$$
where tilde signs denote the Fourier modes labelled by wave vector
$\underline{k}_I$.  This inhomogeneous equation can once again be solved by the
Green function method.  For $k_I \gg k$ the solution is
$$
\tilde \alpha_I (k_I , \tau) \sim {k \cdot k_I\over{k^2_I}} \, {\tilde \rho_I
(k_I)\over \rho} \, , \eqno\eq
$$
which shown that short wavelength fluctuations are
suppressed compared to long wavelength inhomogeneities.

The arguments which lead to (3.17) and (3.19) are self consistent. We could
also reverse the logic and first study the $\alpha$ equation under the
assumption $\rho_I / \rho_0 < 1$, with the result that fluctuations of $\alpha$
on wavelengths smaller than $k^{-1}$ are suppressed.  This result could then be
used in the $\rho$ equation to verify that indeed $\rho_I / \rho_0 < 1$.

All the previous arguments neglect the presence of gauge fields and are thus
only applicable to global symmetries and global defect formation.  We now
demonstrate that the inclusion of gauge fields (working in Coulomb gauge) does
not invalidate the main conclusions.

In the presence of gauge fields, the perturbation equation (3.3) reads
$$
\eqalign{ \partial^2 \rho_I & - \left[  (\partial \alpha_0 - e A_0)^2 + e^2
A^2_0 - 2 \, {\partial V\over{\partial \rho^2}} + 2 {\partial\over{\partial
\rho}} \, {\partial V\over {\partial \rho^2}} \, \rho_0 \right] \rho_I \cr
& = \left[ - g \psi^2 + e^2 A_0 A_I + (\partial \alpha_0 - e A_0) (\partial
\alpha_I - \partial A_I) \right] \rho_0 \, . } \eqno\eq
$$
At the beginning of spinodal decomposition $(T = T_G)$, the amplitude of $A_0
=0$ is constrained by thermal equilibrium considerations.  Note that the
initial configuration at $T_G$ will typically contain magnetic fields with
coherence length $\xi (T_G)$ $^{16)}$.  Hence, it is invalid to set $A_0 = 0$
as initial configuration.  However, since
$$
| D_\mu \psi D^\mu \psi | < T^4 \eqno\eq
$$
and $\psi \sim T$, it follows that $A$ is bounded by
$$
A < e^{-1} T \, . \eqno\eq
$$
Inserting this bound in (3.20) we conclude that the terms involving gauge
fields are negligible provided
$$
T < \lambda^{1/2} \eta  \equiv T_A \, . \eqno\eq
$$
We suspect that improved estimates would show that this ``critical" temperature
is in fact $T_G$.

In the presence of nonvanishing gauge fields, Eq. (3.4) is modified by
$$
\partial^\mu \alpha_0 \rightarrow \partial^\mu \alpha_0 - e A^\mu_0 \eqno\eq
$$
and by the presence of an extra source term $e A^\mu_I \, {\partial_\mu
\rho_0\over \rho_0}$.  However, even at $T = T_G$ the new terms are of the same
order of magnitude or smaller than the ones present in (3.4).  Hence, the
previous conclusions concerning the absence of short wavelength inhomogeneities
in $\alpha_I$ remain true.

In summary, we have shown that even in the presence of gauge fields, the
initial domain structure and winding numbers present at $T = T_A$ are frozen
in.

\chapter{Discussion}

In this paper we have studied the classical dynamics of spinodal decomposition
in the Abelian Higgs model coupled to a thermal bath.  We considered the
classical evolution of an initial field configuration at the Ginsburg
temperature which was assumed to have nontrivial domain structure on a length
scale $\xi (T_G)$  (and hence a finite probability to have nonvanishing winding
number).  We argued that the initial domain structure is stable against thermal
fluctuations.  Hence, the winding number is preserved.  Our results lend
support to the Kibble mechanism for defect formation -- for both global and
local defects.

Above the Ginsburg temperature, thermal fluctuations dominate.  Configurations
with nonvanishing winding number are created and destroyed with a frequency
determined by thermal equilibrium.  Our results indicate that this domain
structure created by thermal fluctuations freezes out at $T_G$, leaving behind
a field configuration with finite probability for nonvanishing winding number
on the scale $\xi (T_G)$.

Our analysis is unfortunately rather qualitative.  It is based on approximate
equations.  We introduced thermal fluctuations by a coupling to an additional
scalar field, but we did not introduce the dissipation term which must be
present in order to be in concordance with the fluctuation-dissipation theorem.
 However, such dissipation terms would only slow the growth of fluctuations and
would hence only strengthen our conclusions.

The analysis of the equations presented in this paper is only approximate.
Only the initial stages of spinodal decomposition in which most nonlinearities
are negligible were analyzed.  Even this analysis was done in an approximate
way.

Our classical analysis can be improved by evolving the solutions of the
equations of motion numerically.  A second improvement would be to study the
Liouville equation associated with our set of classical equations and to derive
a result for $P (\varphi_i, \, \dot \varphi_i , \, t)$, the probability
distribution of the fields$^{15)}$.  Here, $\varphi_i$ stands for the
collection of all fields in the problem.  Finally, the analysis should be
extended to a semiclassical one to control the effects of quantum fluctuations
near the onset of spinodal decomposition.  Work on these projects is in
progress.

\ack

We are grateful to Mark Hindmarsh for many discussions about the issues
discussed here.  We also thank Ed Copeland, Tom Kibble and Ajit Srivastava for
interesting conversations.  This work is supported in part by the US Department
of Energy under contract DE-FG02-91ER40688, Task A, and by an NSF-SERC
collaborative research award NSF-INT-9022895 and SERC GR/G37149.

\REF\one{T.W.B. Kibble, {\it J. Phys.} {\bf A9}, 1387 (1976); \nextline
T.W.B. Kibble, {\it Phys. Rep.} {\bf 67}, 183 (1980).}
\REF\two{R. Brandenberger, {\it Phy. Scripta} {\bf T36}, 114 (1991); \nextline
N. Turok, {\it Phys. Scripta} {\bf T36}, 135 (1991). }
\REF\three{A. Vilenkin and E.P.S. Shellard, `Cosmic Strings and Other
Topological Defects' (Cambridge Univ. Press, Cambridge, 1994).}
\REF\four{T. Prokopec, {\it Phys. Lett.} {\bf B262}, 215 (1991); \nextline
R. Leese and T. Prokopec, {\it Phys. Rev.} {\bf D44}, 3749 (1991).}
\REF\five{T. Vachaspati and A. Vilenkin, {\it Phys. Rev.} {\bf D30}, 2036
(1984); \nextline
A. Albrecht and N. Turok, {\it Phys. Rev. Lett.} {\bf 54}, 1868 (1985).}
\REF\six{J. Preskill, {\it Phys. Rev. Lett.} {\bf 43}, 1365 (1979); \nextline
Ya. B. Zel'dovich and M. Khlopov, {\it Phys. Lett.} {\bf B79}, 239 (1978).}
\REF\seven{N. Turok, {\it Phys. Rev. Lett.} {\bf 63}, 2625 (1989).}
\REF\eight{W. Zurek, {\it Nature} {\bf 317}, 505 (1985).}
\REF\nine{S. Rudaz and A. Srivastava, {\it Mod. Phys. Lett.} {\bf A8}, 1443
(1993).}
\REF\ten{J. Ye and R. Brandenberger, {\it Mod. Phys. Lett.} {\bf A5}, 157
(1990).}
\REF\eleven{I. Chuang, R. Durrer, N. Turok and B. Yurke, {\it Science}, {\bf
251}, 157 (1991); \nextline
B. Yurke, A. Pargellis, I. Chuang and N. Turok, {\it Physica} {\bf B178}, 56
(1992); \nextline
M. Bowick, L. Chandar, E. Schiff and A. Srivastava, {\it Science} {\bf 16}, 943
(1994).}
\REF\twelve{P. Hendry, et al., {\it J. Low Temp. Phys.} {\bf 93}, 1059 (1993).}
\REF\thirteen{M. Hindmarsh, A.-C. Davis and R. Brandenberger, {\it Phys. Rev.
D}, in press (1994).}
\REF\fourteen{J. Langer, {\it Physica} {\bf 73}, 61 (1974).}
\REF\fifteen{R. Brandenberger, H. Feldman and J. MacGibbon, {\it Phys. Rev.}
{\bf D37}, 2071 (1987).}
\REF\sixteen{T. Vachaspati, {\it Phys. Lett.} {\bf B265}, 258 (1991).}
\refout
\end